# *A Classical Probabilistic Computer Model of Consciousness*[*]

By

Stephen Blaha[**]

# ABSTRACT

We show that human consciousness can be modeled as a classical (not quantum) probabilistic computer. A quantum computer representation does not appear to be indicated because no known feature of consciousness depends on Planck's constant h, the telltale sign of quantum phenomena. It is argued that the facets of consciousness are describable by an object-oriented design with dynamically defined classes and objects. A comparison to economic theory is also made. We argue consciousness may also have redundant, protective mechanisms.

# CONTENTS

**This article is an excerpt from the book *Cosmos and Consciousness* by this author. It was felt that this material merited presentation as a paper as well. The book covers additional topics that may be of interest.**

**A View of Consciousness**

There are numerous views of Consciousness. Some of these views attempt to make distinctions between consciousness, the mind, and the brain (body). The mind is the nebulous thing we associate with consciousness, feeling and thought. The body – in particular the brain – is obviously connected to the mind and supports the mind's activity. Yet Consciousness seems endowed with miraculous abilities that many find hard to base entirely on the properties of the brain.

The human brain is in a sense an electromagnetic illusion. *The brain is just as insubstantial as consciousness in reality.*

There is a general lack of appreciation of the power of electromagnetic circuits to create illusions. We see the brain as a hodge-podge of electromagnetic circuitry based on neurons and other brain structures. We then view the mind, and its unity, clarity, powers of logic and analysis, and other features composing one great entity.

It is difficult to reconcile the unity of consciousness of the mind with the brain that implements it. Yet it is more difficult to deny that the mind is based entirely on the brain. Modern research[1] clearly shows the dependence of the properties of the mind on the features of the brain. Consider the effect on the mind of brain diseases or of injuries to the brain.

Modern computer technology actually offers a very clear analogy to the relation of Consciousness and the brain. Consider a modern Personal Computer, a PC. If we open it up we see an ugly hodgepodge of chips and computer circuitry. By only looking at the innards of the PC we have no concept of what this electronic menagerie can generate.

Then we turn on the PC and see the fabulous graphics of a modern computer operating system: lots of windows containing exciting graphics. We can manipulate these windows causing them to change, disappear, reappear with new content, and so on using a mouse, the keyboard or a joystick. We can run captivating multimedia games and simulations with the click of a mouse or the movement of a joystick. We can access and manipulate external information from around the world using the Internet.

Does the computer screen look in any way like the innards of the computer? Does the unity, sophistication and flexibility of the display relate to the odd collection of electronics inside the computer? Obviously not.

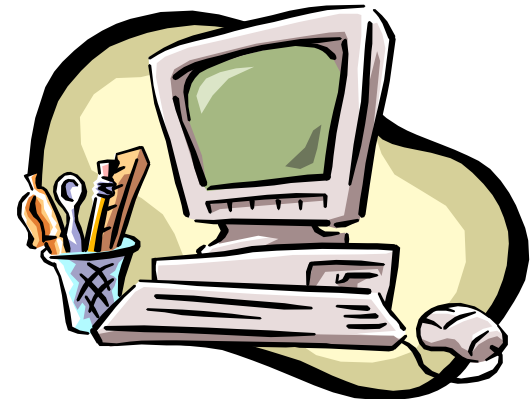

Figure. The ubiquitous PC.

This example is directly analogous to the relation of Consciousness and the brain. The thoughts, unity and activity of Consciousness (the "screen") have no obvious connection to the details of brain (the "computer innards") activity. Yet the mind is a construct of the electrical activity in the brain.

---

[1] Gerald M. Edelman and Giulio Tononi, *A Universe of Consciousness*, (Basic Books, New York, 2000). There are many other excellent books on consciousness. See the references in Edelman and Tononi or search the Web.

The Consciousness of the mind is the combined result of the electrical activity of the brain.

**Consciousness: Quantum or Classical Probabilistic**
Our studies of space, time, and matter – the Cosmos – have led us to nothingness. Consciousness itself is not material. It is also nothingness. Both Consciousness and the Cosmos are given shape by laws. The laws structure the "nothingness" and provide the "nothingness" with features and properties.
In the case of the Cosmos we have made a case for a Quantum Computer formulation of the fundamental theories of Physics.

In the case of Consciousness we propose that Consciousness be best viewed within the framework of Classical Probabilistic Computers.

A Classical Probabilistic Computer is a purely classical computer (no quantum effects) that produces a variety of different outputs from a given input to the computer. Each possible output has a certain probability of occurring. The probabilities are all strictly classical – they are not of quantum mechanical origin.

A Classical Probabilistic Computer can be viewed as:

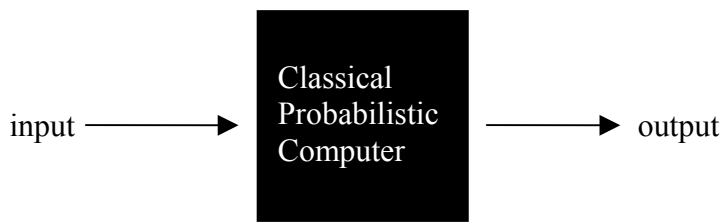

Figure. A Classical Probabilistic Computer produces one output from a given input. The output is one of a number of possibilities.

If the same program is run over and over again in a Classical Probabilistic computer then a variety of outputs will occur. Each output will appear in the set of outputs with a frequency proportional to its probability of occurrence.

The reason for suggesting that Consciousness be modeled as a Classical Probabilistic Computer is based on the following thoughts:

1. Consciousness appears to be a classical phenomenon. If we consider the properties of the mind there is no convincing evidence for significant quantum effects. Even if Science should find isolated quantum phenomena surfacing in experiments on Consciousness the overwhelming bulk of the phenomena of Consciousness is still not quantum but classical in nature.

2. Conscious activity evolves in time through a series of states. At any given moment Consciousness has billions upon billions of states that it can evolve into (see reference 32 for graphic descriptions of the time evolution of conscious states). Given this vast number of possible states we must treat the evolution of consciousness with time as a statistical probabilistic phenomenon.

So we conclude that we must treat Consciousness as a classical, probabilistic phenomena in principle.

**The Problem of Consciousness – The Lesson of the Conch**
After determining that Consciousness is classical physics and chemistry and best treated as a statistical probabilistic phenomenon we confront the overwhelming complexity of Consciousness.

We also confront Nature's protective mechanisms that may obscure our understanding of Consciousness. Consider the conch *Strombas gigas*.

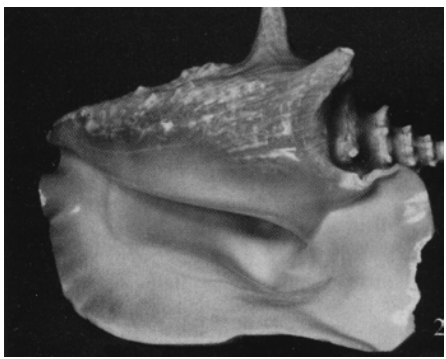

Figure. The conch *Strombas gigas*.

Ninety-nine per cent of this giant pink conch is made of a mineral called aragonite that is a form of calcium carbonate that breaks like chalk. Yet the shell of the conch resists fractures a hundred to a thousand times better than the mineral of which it is formed. Nature has developed a microscopic structure for the conch that surrounds each aragonite crystal in its shell with a protein that changes the toughness of the shell by enabling fractures to spread without breaking the material. In addition the shell has three layers with the "grain" of each layer perpendicular to the grain of adjacent layers. This composite cross-grained material gives the conch shell extraordinary strength.

If Nature expended such effort during evolution to protect the humble conch, then what effort must have been expended to protect the workings of the Consciousness of Man?

Coincidentally the brain has three main neuroanatomical arrangements. First there is the thalamocortical system that networks the thalamus, the cortex and cortical regions. Secondly, there is a network of long polysynaptic loops that extend between the cortex and the cortical appendages. Thirdly, there is the diffuse network of projecting value systems (the noradrenergic locus coeruleus) that extends over the entire brain. The projecting value systems network appears to fire (react) whenever an important event happens such as a loud noise. When it fires it causes the release of neuromodulator chemicals that appear to influence the resulting neural response to the event. The projecting value system may be a way of protecting the brain against over-reacting to major disturbing events.

**The Current Theory of Consciousness**
Realizing the complexity of the phenomena of Consciousness and the added complexity of protective mechanisms that Nature might have built into the structure of Consciousness it is no surprise that we do not have a satisfactory Theory of Consciousness.

This situation is not without precedent. Similar situations have occurred in the "hard" sciences and in the social sciences. For example, George Uhlenbeck, the co-discoverer of electron spin and one of the outstanding physicists of the mid-twentieth century, spent many years trying to develop a satisfactory theoretical framework for understanding Statistical Mechanics from a microscopic point of view. He told this author (about 1970) that he felt he did not succeed. Uhlenbeck had the advantage of a completely known theory of microscopic particles and a well-known theory of the Statistical Mechanics of large numbers of particles. Despite these advantages he was not able to relate the microscopic theory with the theory of the Statistical Mechanics of a large number of microscopic particles. Relating different levels of theories such as a microscopic theory and a macroscopic theory is difficult.

The situation of theories of Consciousness and theories of the brain is much less favorable. We know the overall neuroanatomy (structure) of the brain. We have a pretty good idea of how some features such as vision map to specific brain areas. We have a decent understanding of brain neurochemistry. We have a lot of data on features of Consciousness and some ideas on how these features map to brain features. But we do not have a detailed understanding of the brain. And we do not have a complete understanding of Consciousness. In particular we can usually only make qualitative statements about Consciousness. We don't even know what the relevant variables are for conscious phenomena. Who can say how to quantify emotions such as fear or anger? We need at least a Richter scale for emotion.

Given this state of affairs a detailed theory of Consciousness similar to a theory of Physics or Chemistry is no where in sight. We can only expect qualitative descriptions and rules for most phenomena of Consciousness. We can only expect general relationships between brain activity and phenomena of Consciousness. We can expect certain specialized (simple) phenomena of Consciousness to be based on detailed brain activity.

**A Similarity between Theories of Consciousness and Economic Theory**

The study of Consciousness is plagued by the lack of a quantitative framework to describe phenomena. We don't know the relevant variables that describe a phenomenon of Consciousness. We usually don't know what to measure, and, in the cases where we do find something to measure, we don't know how to measure it or how to interpret it or how to relate it to brain activity quantitatively.

This state of affairs is reminiscent of the situation of the Economics of a country. At the microscopic level we can in principle trace every transaction, aggregate all the transactions in the country's economy and thus obtain a complete view of the economy. We can also trace the evolution of the economy in time. However we do not have a detailed complete quantitative theory of Economics.

As a result we can only make predictions based on extrapolations of trends. If we change the pattern of financial transactions in the country we cannot unambiguously predict the effects on the economy. We can only create models based on assumptions. Some models are quite good. But they are no replacement for a complete theory of Economics.

The modern theory of Economics was born in the Eighteenth and Nineteenth Centuries in the work of Adam Smith and others. It started with general qualitative statements based on simple observations. These statements had some predictive power. Then in the Twentieth Century a host of Economists developed quantitative theories for economic phenomena. Economics became semi-quantitative – but there were still many unanswered questions. There is still a problem relating the microscopic picture of individual transactions and the "Big Picture" view of the economy. The predictive power of Economic Theory is still spotty.

Compare the development of Economic Theory with the Theory of Consciousness. The microscopic theory is the theory of the brain. The "Big Picture" is Consciousness. We can only make quantitative statements about Consciousness. Our microscopic picture is still incomplete. Clearly the state the development of Consciousness Theory is comparable to the state of Economic Theory in the Nineteenth Century. On the positive side the rate at which our knowledge of Consciousness and the brain is developing is much faster than the development of Economics.

The development of Economics offers a paradigm for the development of our understanding of Consciousness.

It also suggests a way of picturing the relation between Consciousness and the brain. We can view the brain as a vast interconnected network of electrical activity with connections to the Consciousness. We will view Consciousness as a separate level that is conceptually unified and connected by "channels" or communications paths to the brain. This is a theoretical framework that reflects strategies used in economic analysis. One can view stock or commodity prices from two perspectives: one perspective views price changes as reactions to external events; another perspective views price changes from a "technical" perspective based on trends in charts of historical price data.

We suggest that one should view Consciousness as a thing in itself developing a self-contained theory of Consciousness (a "technical" approach). This theory can then be related to the underlying dynamics and processes of the brain.

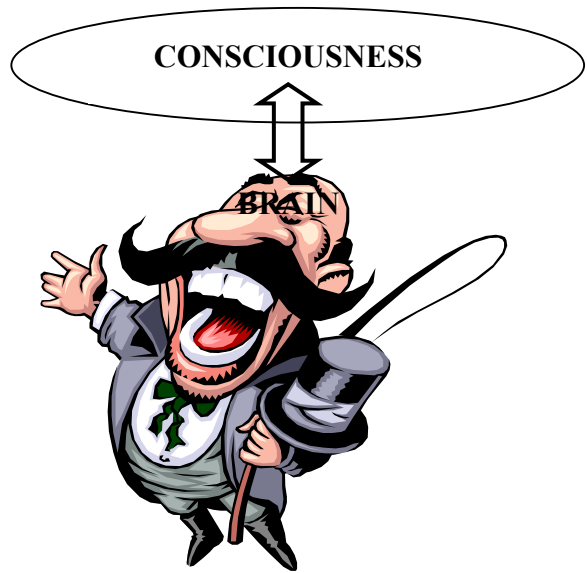

Figure. Consciousness as a separate conceptual entity.

By separating Consciousness from the brain and establishing a structured interface between the brain and Consciousness one can hope to develop a provisional model of Consciousness.

**A Probabilistic Computer (PC) Model of Consciousness**
Although human Consciousness is large and complex it must be finite since it is derived from the human brain which is finite. We have seen that Consciousness is overwhelmingly classical (not quantum) in nature and that it must be treated probabilistically because of the billions upon billions of states of Consciousness.

These considerations lead us to suggest a Classical Probabilistic Computer Model for Consciousness with strict interfaces to the human brain. The human brain is a source of inputs and outputs for the computer.

Many writers argue that human Consciousness cannot be described by a computer model (see reference 32 for an example). Then they sometimes proceed to use a computer model to simulate some feature of Consciousness. Since human Consciousness is based on a finite human brain that in principle can be simulated by a sufficiently large and complex computer it seems reasonable to think that Consciousness can be modeled on a computer with appropriate features and capabilities.

Does Consciousness "run" like a computer program? No, a better view of Consciousness is to view it as a set of capabilities and features that interconnect to constitute Consciousness. Each of these entities (They may map to groups of neurons in the brain.) has a set of capabilities or features.

One can think of each of them as an "object" that has a specific set of capabilities and features. These objects have a "mini-program" inside them that specifies their behavior and how they hook up with other objects to perform tasks and to constitute the Consciousness. The hooks are variable and dynamic.

The time evolution of the Consciousness from state to state is a result of the execution of these "mini-programs" in a dynamic ever-changing way. There is no overall program but instead there is an ever changing, dynamic, unfolding of states of Consciousness in response to external inputs and based on the previous state of Consciousness plus random effects within Consciousness.

The description of Consciousness as a collection of objects with features, an internal "mini-program" describing the object's behavior and interaction with other objects can be called an Object-Oriented definition. The concept of Object-Oriented Programming is currently the preferred way to program amongst computer software developers. It meshes well with the observed features of Consciousness. One major difference is the dynamic nature of the grouping of neurons in response to external inputs. The Object-Oriented programming parallel would be to have class definitions dynamically reforming in response to the evolution of a program.

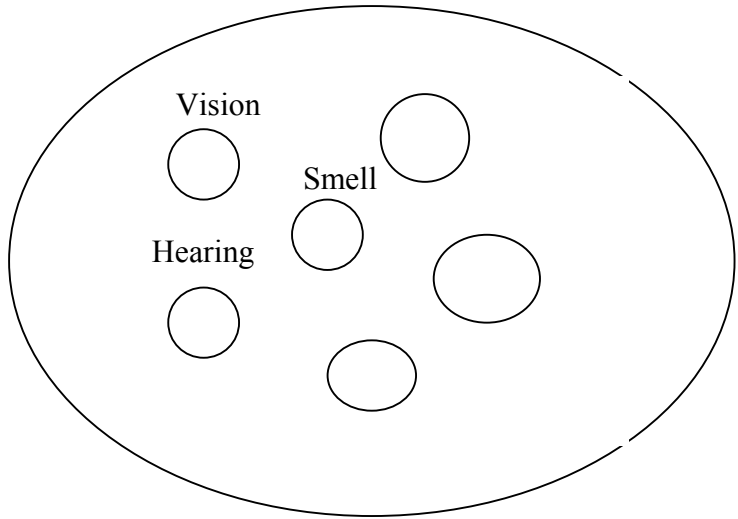

Figure. An Object-Oriented view of Consciousness.

Thus Consciousness does not have a program in the old fashioned sense of the word. It has a dynamically changing, event driven program with program fragments in each of the parts of Consciousness. These program fragments can dynamically link together in response to events to take Consciousness from one state to another. The pattern of linkings is driven by a complex set of interconnections between the parts of Consciousness. Dynamic Linking is the preferred way of creating a computer program (a .EXE program) in modern computing.

There are two major features of Consciousness that are of crucial importance in defining a computer representation of Consciousness at this level of discussion:

1. Consciousness can marshal its resources to allocate more resources to important tasks that it faces.

2. Consciousness can be "rewired" to adapt to meet short-term needs and to meet long-term needs.

These features tell us about the computer mechanisms or mini-programs driving the time development of Consciousness (the way in which a Consciousness evolves from state to state in time).

The first point tells us that a correct Probabilistic Computer model of Consciousness must be able to dynamically reallocate its resources in response to inputs.

The second point tells us that the mini-programs that describe the changes in time of the state of Consciousness must be able to change itself. The LISP programming language is an example of a language whose programs are capable of changing themselves as they execute (evolve). Another way of implementing this feature at the hardware level is to say the computer can rewire itself to meet new needs. In fact this process is known to happen in the human brain. Human brains can rewire themselves over time if a brain is damaged or in response to stimuli.

These additional features make the Probabilistic Computer a suitable model for Consciousness.